\documentclass[11pt]{article}
\usepackage{moriond,epsfig}

\def\be{\begin{equation}}
\def\ee{\end{equation}}
\def\bea{\begin{eqnarray}}
\def\eea{\end{eqnarray}}

\usepackage{amsmath}
\usepackage{amssymb}
\newcommand{\Ca}{\ensuremath{C_{\!A}}}
\newcommand{\Cf}{\ensuremath{C_{\!F}}}
\newcommand{\Nc}{\ensuremath{N_{\!C}}}

\newcommand{\gs}{\ensuremath{g_s}}
\begin{document}
\begin{flushright}
  CERN-PH-TH/2010-133
\end{flushright}
\title{$CP$-properties of the Higgs-boson couplings from $H$\ +\ dijets through
  gluon fusion}

\author{Jeppe R. Andersen}

\address{Theory Division, Physics Department, CERN, CH-1211 Geneva 23, Switzerland}
\boldmath
\maketitle\unboldmath\abstracts{At lowest order in perturbation theory, the production
  of a Higgs boson in association with dijets displays a strong correlation
  in the azimuthal angle between the dijets, induced by the $CP$-properties
  of the Higgs Boson coupling. However, the phase space cuts necessary for a
  clean extraction of the $CP$-properties in the gluon fusion channel
  simultaneously induce large corrections from emissions of hard radiation
  and thus formation of additional jets. This contribution discusses how the
  $CP$-properties of the Higgs boson coupling can be cleanly extracted from
  events with more than two jets, based on a technique developed from insight
  into the high energy limit of hard scattering matrix elements.}

\section{Introduction}
\label{sec:introduction}

One of the primary goals of experiments at the CERN 
Large Hadron Collider (LHC) is the search for the Higgs boson(s) 
which, within the Standard Model (SM) and many of its extensions,
provide direct access to the dynamics of electroweak symmetry breaking.
Once discovered, the focus of Higgs physics will turn to the study of 
Higgs boson properties, like its mass, spin, \ensuremath{CP} parity and the strength and
structure of Higgs boson couplings to heavy fermions and gauge bosons.

Among the various Higgs channels at the LHC, the production of a Higgs boson
in association with two energetic jets has emerged as particularly promising
in providing information on the dynamics of the Higgs sector. This is true in
particular for the gluon fusion channel, where the $CP$-properties of the
Higgs boson couplings to the fermions in the loop-induced coupling can be
extracted\cite{Klamke:2007cu}: Tree-level considerations lead to the
expectations of a strong azimuthal correlation between the two jets, with a
phase depending on the relative weight of a $CP$-even (SM-like) and $CP$-odd
coupling. The azimuthal angle modulations get particularly pronounced when
the two jets are widely separated in rapidity. Equivalent effects are
expected in vector boson fusion and have been discussed
in Ref.\cite{Plehn:2001nj,Hankele:2006ma} for the idealised situation of parton
level events at leading order. 

The extraction of the $CP$-properties of the Higgs boson couplings in gluon
fusion will require some cut on the rapidity separation between the two hard
(e.g.~$p_\perp>40$GeV) jets; typically, they are required to be at least 3
units apart in rapidity\cite{Klamke:2007cu}, or
alternatively\cite{Andersen:2010zx} the Higgs boson is required to be
produced between the dijets in rapidity, with a minimum distance of .5-1
units of rapidity between the Higgs boson and the hard jets.

\section{Hard Radiative Corrections}
\label{sec:beyond-tree-level}

The tree-level observations leading to the expectation of the azimuthal
correlation are jeopardised by the requirement of a size-able rapidity
separation between the jets. For the gluon fusion channel, this requirement
increases the hard radiative corrections leading to the formation of
additional jets; and therefore one must address the problem of how to extract
the $CP$-properties of the Higgs boson couplings from events with strictly
more than two jets, where one might think it is not so clear how to
generalise the azimuthal angle studied for events of pure Higgs-boson plus
dijet. It is clear that the study of just the azimuthal angle between any two
jets (e.g.~the two hardest) will necessarily be less correlated once real
radiative corrections are taken into account. This contribution discusses how
to form an observable, so that the extraction of the $CP$-properties is
stable against radiative corrections\cite{Andersen:2010zx}.

\begin{figure}
  \centering
  \epsfig{width=.45\textwidth,file=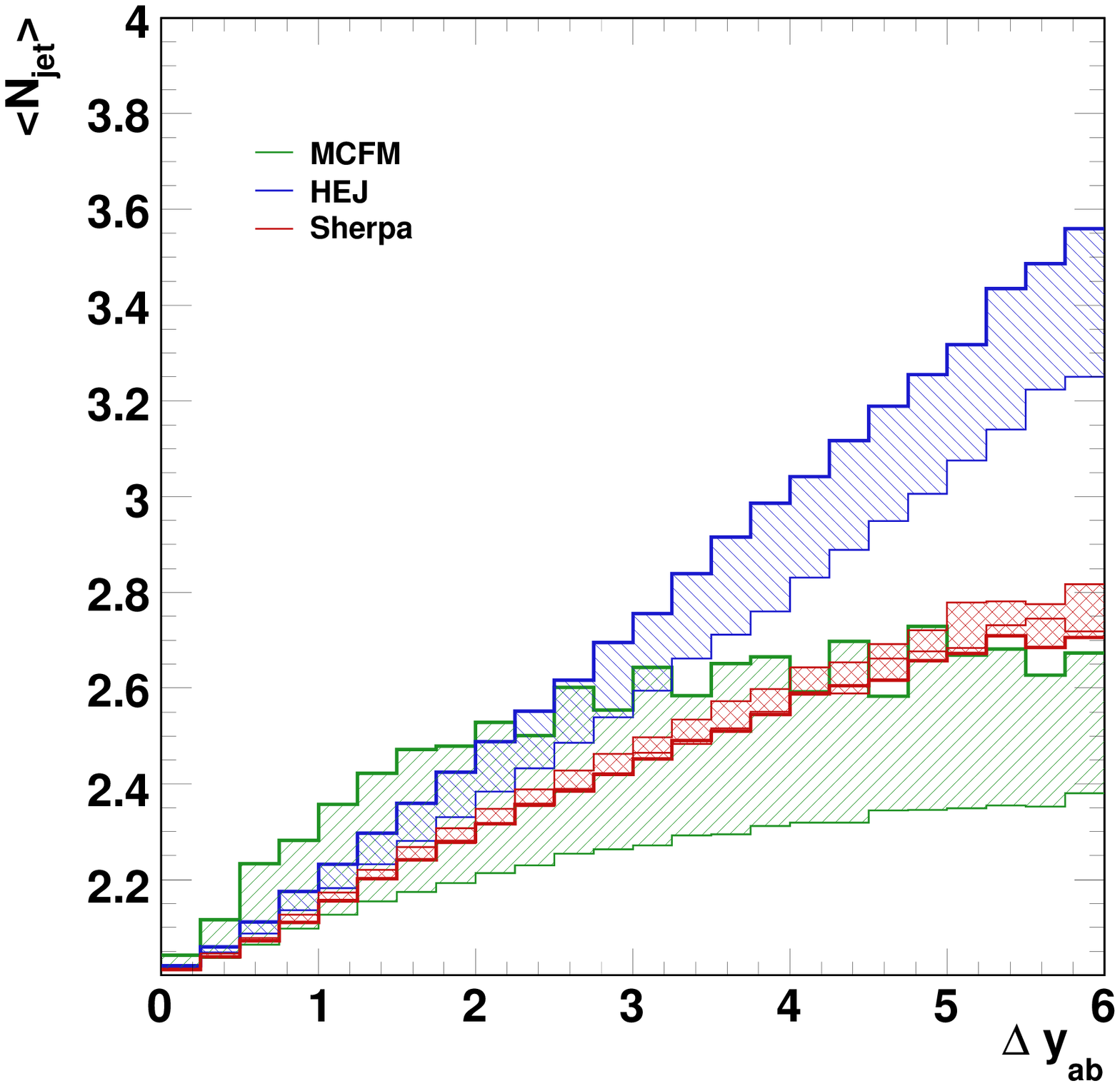}\hfill
  \epsfig{width=.45\textwidth,file=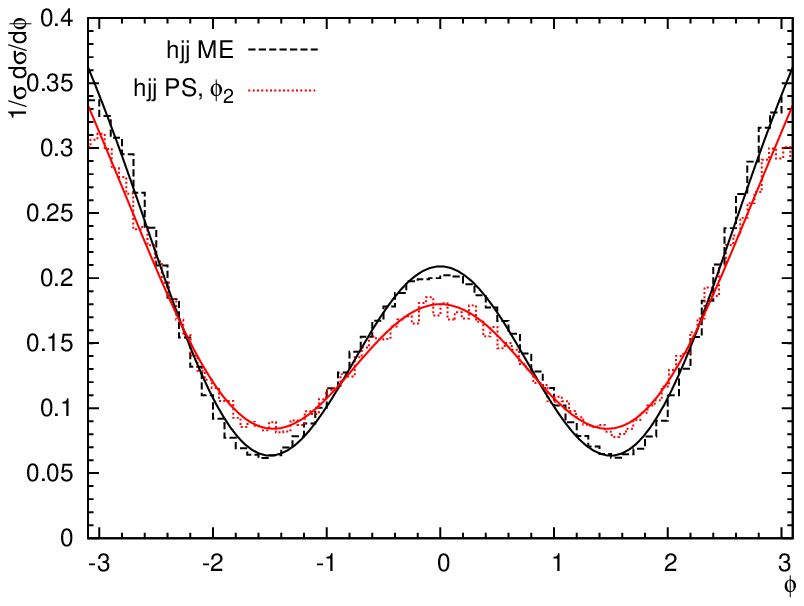}
  \caption{Left: The average number of hard jets ($p_\perp>40$GeV) in
    Higgs-boson production in association with dijets as a function of the
    rapidity difference between the most forward and most backward hard jet. Figure taken from
    Ref.$^6$ Right: The differential cross section on the azimuthal angle
    discussed in the text for the tree-level calculation (black) and for
    showered and hadronised events (red). Figure taken from Ref.$^6$}
  \label{fig:HAvgJets}
\end{figure}

First, we will briefly discuss the reason for the increasing weight of real,
hard radiative corrections as the rapidity span between the dijets is
increased. This is caused by two effects. First, two widely separated (in
rapidity) jets will dominate the contribution to the light-cone momentum
fraction of the partons extracted from the proton, so the relative impact of
extracting a little extra energy from the proton in order to form an
additional central jet is small (the details will obviously depend on just
how steeply the parton density functions are falling with increasing
$x$). Secondly, the phase space for the emission of additional radiation
increases as the rapidity span between the most forward and most backward jet
is increased. These kinematic considerations are shared of course by all
processes, and by all models for the these (e.g.~shower MC, fixed order
perturbation theory). The amount of hard radiation generated obviously
differs between different processes (e.g.~Higgs boson+dijets through weak
boson fusion or gluon fusion\cite{Dokshitzer:1991he}), and between different
models of a given process (e.g.~shower MC, NLO, resummation). To illustrate
this last effect, Figure~\ref{fig:HAvgJets} (taken from
Ref.\cite{Binoth:2010ra}) displays the average number of jets 
in events (at a $pp$-machine with $\sqrt{s}=10$TeV) with a Higgs-boson in
association with at least two hard jets (of transverse momentum greater than
40GeV) as a function of the rapidity span between the most forward and most
backward hard jet, as calculated at fixed next-to-leading
order\cite{Campbell:2010cz} (green), \texttt{Sherpa}\cite{Gleisberg:2008ta}
with tree-level matching up to Higgs-boson plus four partons using
\texttt{Comix}\cite{Gleisberg:2008fv} (red), and finally an all-order sum of
the leading radiative corrections for widely separated
emissions\cite{Andersen:2008ue,Andersen:2008gc,Andersen:2009nu,Andersen:2009he}
(blue). The width of the bands indicate the scale variation, but the initial
choice is different and the range of variation is smaller in \texttt{Sherpa}
than in the two other models. We see that all models for this process
predicts a strong correlation between the rapidity span between the most
forward and most backward hard jet, and the average number of hard jets (all
above 40GeV in transverse momentum) in the event. In fact, the increasing
relevance of the high-multiplicity states with growing rapidity span is a
central motivation for the BFKL\cite{Kuraev:1977fs,Balitsky:1978ic}
resummation programme for hard processes. Indeed, the strong correlation
between the rapidity span of the event and the average number of hard jets
were observed in variants\cite{Orr:1997im,Andersen:2001kta} on the BFKL
formalism also for pure jets\cite{Andersen:2003gs} and
W+dijets\cite{Andersen:2001ja}. While the BFKL formalism reproduces the limit
of the full QCD amplitudes for infinite rapidity separation between all
(hard) particles, the formalism developed in
Ref.\cite{Andersen:2008ue,Andersen:2008gc,Andersen:2009nu,Andersen:2009he}
obeys also other constraints (e.g.~gauge-invariance) in all of phase space
(i.e.~also for sub-leading kinematics).

Figure~\ref{fig:HAvgJets} also indicates that for the rapidity spans of
interest for the extraction of the $CP$-properties, the average number of
jets is significantly larger than 2. For the NLO calculation, the exclusive
2-jet and 3-jet rates have to be equal, in order to get an average number of
hard jets of 2.5. It is clear that understanding the pattern of multi-jet
radiation will be important for a stable extraction of the $CP$-properties of
the Higgs-boson couplings.

\vspace{-2mm}
\section{Lessons From The High Energy Limit}
\label{sec:high-energy-limit}
In order to generalise the lowest order study of the azimuthal angle between
the dijets to the case of multiple hard jets we start by studying the (colour
and helicity summed and averaged) square of the matrix element for $gg\to
g\cdots ghg \cdots g$ in the limit of infinite rapidity separation between
each produced particle (the so-called multi-Regge-kinematic (MRK) limit):
\begin{align}
  \label{eq:ngluonplush}
  \left| \overline{\mathcal{M}}_{gg\to g\cdots ghg\cdots g}\right |^2 \rightarrow \frac {4 {\hat s} ^2}
    {\Nc^2-1}\ 
   \left( \prod_{i=1}^j\frac{\Ca\ \gs^2}{\mathbf{p}_{i\perp}^2} \right)
   \frac{|C^H(\mathbf{q}_{a\perp}, \mathbf{q}_{b\perp})|^2}{\mathbf{q}_{a\perp}^2\  \mathbf{q}_{b\perp}^2}\
   \left(\prod_{i=j+1}^n\frac{\Ca\ \gs^2}{\mathbf{p}_{i\perp}^2}\right),
\end{align}
where $\mathbf{q}_{a\perp}=-\sum_{i=1}^j\ \mathbf{p}_{i\perp}$, where $j$ is the number of gluons
with rapidity smaller than that of the Higgs boson, and
$\mathbf{q}_{b\perp}=\mathbf{q}_{a\perp}-\mathbf{p}_{h\perp}$. In this limit, the contribution from
quark-initiated processes is found by just a change of one colour factor
$\Ca\to\Cf$ for each incoming gluon replaced by a quark. The effective
vertex for the coupling of a SM Higgs boson to two off-shell gluons through a
top loop is in the combined large-$m_t$ and MRK limit\cite{DelDuca:2003ba}
\begin{align}
  \begin{split}
    \label{eq:Ch}
    C^H(\mathbf{q}_{a\perp},\mathbf{q}_{b\perp})\ &=\ i\ \frac A 2\ \left(|\mathbf{p}_{h\perp}|^2 -
      |\mathbf{q}_{a\perp}|^2 - |\mathbf{q}_{b\perp}|^2 \right)=-iA\ \mathbf{q}_{a\perp}\cdot\mathbf{q}_{b\perp},\\
    A\ &= \frac {\alpha_s} {3\pi v},\quad v=246\ \mbox{GeV}.
  \end{split}
\end{align}
In the simple case of $hjj$ at tree level in the SM we recover from
Eq.~(\ref{eq:Ch}) a cosine modulation in the azimuthal angle between the two
jets, which is indeed the correct limiting behaviour seen in the full
tree-level matrix element. A CP-odd contribution to the coupling would
introduce a sinus-component, and a phase-shift in the angular distributions
discussed later. However, Eq.~(\ref{eq:Ch}) also hints how to
recover this azimuthal modulation in events with more than two
jets\cite{Andersen:2010zx}: simply divide the jets into two sets according to
whether their rapidities are smaller or greater than that of the Higgs
boson; then calculate the azimuthal angle between the transverse sum of
vectors from each set. This angle will in the MRK limit display the same
behaviour as that of the azimuthal angle between the two partons in the
lowest order analysis.

\section{Results}
\label{sec:results}
In Ref.\cite{Andersen:2010zx} we checked the stability of the angle as
defined above against several corrections beyond the tree-level description,
and will here present just a few of the findings. The first thing one could
worry about is the stability against the effects, both perturbative and
non-perturbative, included in a general-purpose Monte Carlo generator. In
Fig.~\ref{fig:HAvgJets}~(right) we compare the azimuthal modulation using the
definition discussed in the previous section found at tree-level with that
found after showering and hadronisation of these states with
\textsc{Herwig{\footnotesize++}}\cite{Bahr:2008pv}. We see that the azimuthal
modulation survives the effects of hadronisation etc., and also that the
real emission from the shower, which does not end up in hard jets (and is thus not
included in the construction of the azimuthal angle), does not spoil the
positions of the peaks and troughs of the distribution. 

\begin{figure}
  \centering
  \epsfig{width=.45\textwidth,file=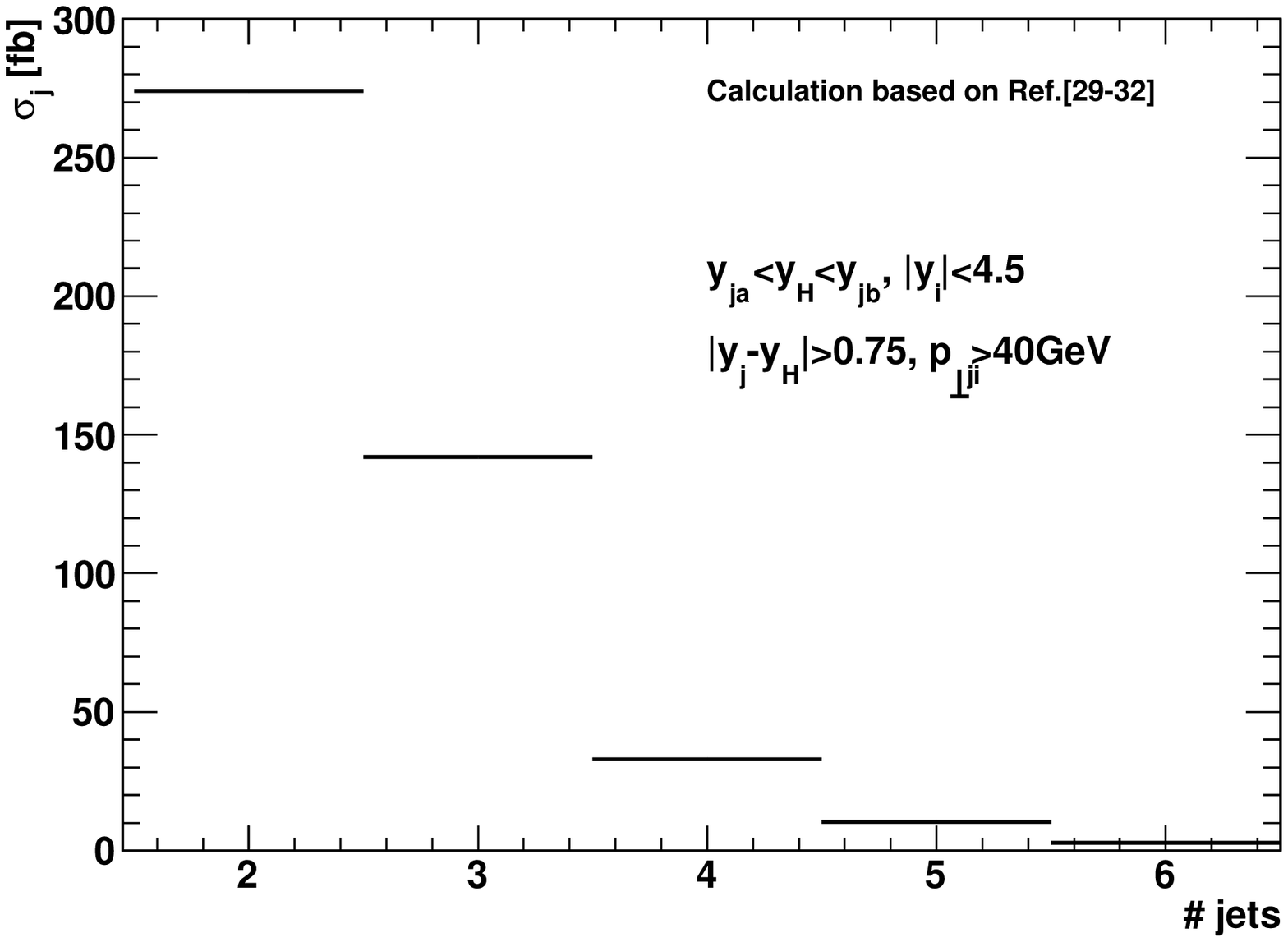}\hfill
  \epsfig{width=.45\textwidth,file=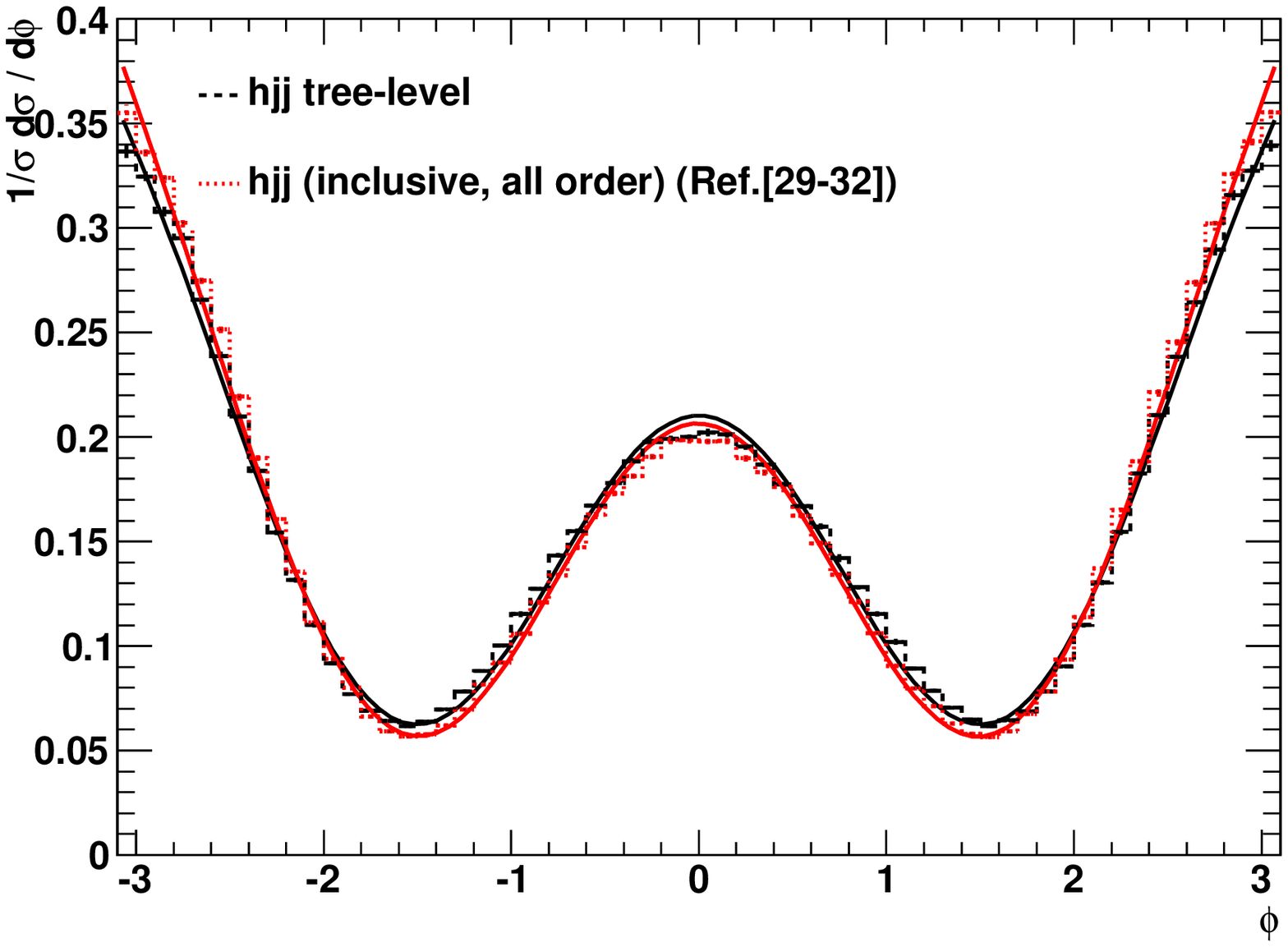}
  \caption{Left: The jet rates in the inclusive all-order sample generated with the
    method discussed in Ref.$^{10,11,12,13}$. The exclusive 2-jet rate
    accounts for only roughly 60\% of the inclusive 2-jet sample. Right: The
    distribution on the azimuthal angle discussed in the text as
    calculated at tree-level (black) and in the all-order sample. Figure taken from Ref.$^4$}
  \label{fig:HEJ}
\end{figure}

While the shower-formalism correctly resums the soft- and collinear radiation
from the tree-level $hjj$-configuration, the pure shower-formalism
underestimates the amount of hard radiation, which can lead to further
decorrelation. In order to check the stability of the azimuthal distribution,
against such radiation, we analyse the constructed azimuthal observable on a
set of $hjj$-events generated in the all-order formalism discussed
earlier\cite{Andersen:2008ue,Andersen:2008gc,Andersen:2009nu,Andersen:2009he}. In
Fig.~\ref{fig:HEJ} we show on the left the distribution of the number of hard
jets in the event sample within the cuts mentioned in the figure. The
exclusive 2-jet rate accounts for around 60\% of the inclusive two-jet rate,
so it is clearly necessary with a strategy for a stable extraction of the
$CP$-properties of the Higgs boson couplings for events with strictly more
than two jets. In Fig.~\ref{fig:HEJ} (right) we have used the same event
sample as used for the plot on the left, and show the differential
distribution on the azimuthal angle constructed as discussed. Furthermore, we
compare this to the result obtained at lowest order. 

In conclusion, the constructed azimuthal observable is clearly very stable
against higher order perturbative corrections, allowing for a stable
extraction of the $CP$-properties of the Higgs boson couplings in gluon
fusion.

\vspace{-5mm}
\section*{References}
\label{sec:references}

\end{document}